# Video and Synthetic MRI Pre-training of 3D Vision Architectures for Neuroimage Analysis

Nikhil J. Dhinagar*[1], Amit Singh[1], Saket Ozarkar[1], Ketaki Buwa[1], Sophia I. Thomopoulos[1], Conor Owens-Walton[1], Emily Laltoo[1], Yao-Liang Chen[2], Philip Cook[3], Corey McMillan[3], Chih-Chien Tsai[4], J-J Wang[5], Yih-Ru Wu[6], Paul M. Thompson[1]

[1]Imaging Genetics Center, Mark & Mary Stevens Neuroimaging & Informatics Institute, Keck School of Medicine, University of Southern California, Los Angeles, CA, USA
[2]Department of Diagnostic Radiology, Chang Gung Memorial Hospital, Keelung, Taiwan
[3]Department of Neurology, Perelman School of Medicine, University of Pennsylvania, Philadelphia, PA, USA
[4]Healthy Aging Research Center, Chang Gung University, Taoyuan, Taiwan
[5]Department of Medical Imaging and Radiological Sciences, Chang Gung University, Taoyuan, Taiwan
[6]Department of Neurology, Chang Gung Memorial Hospital, Linkou, Taoyuan, Taiwan

**Abstract.** Transfer learning represents a recent paradigm shift in the way we build artificial intelligence (AI) systems. In contrast to training task-specific models, transfer learning involves pre-training deep learning models on a large corpus of data and minimally fine-tuning them for adaptation to specific tasks. Even so, for 3D medical imaging tasks, we do not know if it is best to pre-train models on natural images, medical images, or even synthetically generated MRI scans or video data. To evaluate these alternatives, here we benchmarked vision transformers (ViTs) and convolutional neural networks (CNNs), initialized with varied upstream pre-training approaches. These methods were then adapted to three unique downstream neuroimaging tasks with a range of difficulty: Alzheimer's disease (AD) and Parkinson's disease (PD) classification, "brain age" prediction. Experimental tests led to the following key observations: 1. Pre-training improved performance across all tasks including a boost of 7.4% for AD classification and 4.6% for PD classification for the ViT and 19.1% for PD classification and reduction in brain age prediction error by 1.26 years for CNNs, 2. Pre-training on large-scale video or synthetic MRI data boosted performance of ViTs, 3. CNNs were robust in limited-data settings, and in-domain pretraining enhanced their performances, 4. Pre-training improved generalization to out-of-distribution datasets and sites. Overall, we benchmarked different vision architectures, revealing the value of pre-training them with emerging datasets for model initialization. The resulting pre-trained models can be adapted to a range of downstream neuroimaging tasks, even when training data for the target task is limited.

**Keywords:** Transfer learning, vision transformers, MRI.

# 1   Introduction

Magnetic resonance imaging (MRI) is widely collected globally for clinical evaluation of neurological and psychiatric conditions, as well as for neuroscience research. AI methods can assist in processing these scans and, with increasing automation, may ultimately lead to faster and more accurate diagnosis and prognosis. In the U.S., for example, the Food and Drug Administration (FDA) has over the last few years reviewed and approved devices with AI and machine learning (ML) capability [23].

Transfer learning is an approach to deep learning where knowledge from a task that has already been solved - often called the "pre-training" task - is transferred to a novel task. As vast datasets of 2D natural images are publicly available, such as ImageNet [3], they have often been used to pre-train image labeling and classification methods, even for 2D or 3D medical images [1][18].

In contrast, 3D medical image analysis has remained a unique domain for deep learning given the challenges of limited availability of expertly labeled data, problems relating to overfitting where AI models may not generalize well to data from new sites and scanners, and model architectures that can learn 3D spatial contextual features. Willemink et al. discussed the benefits of foundation models [25] and their adaptation towards more specialized medical tasks. Lu et al. [14] showed that sex classification of 3D brain MRI at a large scale was a pre-training strategy for other tasks, such as predicting a person's age from their MRI (brain age prediction). Dufumier et al. proposed a modified contrastive loss [2] [7] to incorporate continuous meta-data during pre-training. More recently, synthetically generated brain MRI [17] data and even natural video clips [10] [15] are new methods to pre-train 3D models [19] [11], to natively learn generic 3D features from large-scale datasets. Recent efforts have focused on adapting attention-based vision transformers (ViTs)[5] to the medical imaging domain [16] [20]. Masked image modeling has made a resurgence as an effective self-supervised learning (SSL) technique [8] [26] using a transformer-based encoder-decoder architecture.

In this work, we benchmark ViT and CNN-based vision architectures that use 3D T1-weighted (T1-w) brain MRI scans as inputs and leverage a range of pre-trained initializations for neuroimaging tasks. To test our models on downstream tasks of varying difficulty, we focus on Alzheimer's disease (AD) and Parkinson's disease (PD) classification - two common tasks that use brain MRI to detect specific neurodegenerative diseases - and brain age prediction. Age prediction from MRI is commonly studied as a benchmarking challenge and has been widely used to yield a biomarker of "brain age" to quantify deviations from healthy aging. Although prior works have achieved good performance for AD classification, PD classification is challenging using T1-w MRI scans.

The novel contributions of this work include four key take-aways:

1. Pre-training improved performance across three unique neuroimaging tasks - AD and PD classification, and brain age prediction.

2. Emerging large-scale pre-training datasets created with video clips and synthetic MRI scans boosted the performance of ViT-based architectures.
3. CNNs achieved robust performance in limited data scenarios and achieved a boost from pre-trained in-domain initialization.
4. Pre-training improved generalization to new datasets, sites, and tasks.

## 2 Methods

### 2.1 Vision Architectures

For our experiments, we chose five recent 3D models with a range of sizes, from 1.8 million (M) to 27M parameters. This included: the **SwinT** vision transformer with 27M parameters, as adapted for video analysis [13], the multi-instance neuroimage transformer (**MiNiT**) with 3.6M parameters, and the Neuroimage Transformer (**NiT**) with 2M parameters, recently adapted for neuroimage applications [20]. Further, we used a commonly used architecture in neuroimaging - the **DenseNet121** convolutional neural network (CNN) [9] with 11M parameters, and a scaled down version with 10 times fewer parameters, **Tiny-DenseNet CNN** [6] with 1.8M parameters. We scaled-down the **ViT** B/16 [5] with 12 attention heads, 6 encoder layers and coupled it with a lightweight decoder for our masked image modeling experiments. The patch sizes, MLP dimension, and hidden dimension used for each ViT was left unchanged from the original design.

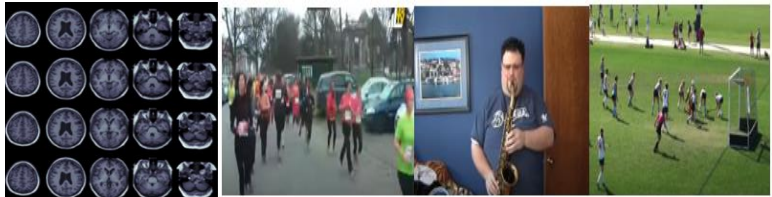

**Figure 1.** Synthetic and out-of-domain pre-training data. Left - T1-w brain MRI generated from the LDM dataset, Right - Snapshots of video clips from Kinetics-400 with labels "jogging", "playing saxophone", "playing field hockey".

### 2.2 Pre-Training Approaches and Target Fine-tuning Tasks:

The methods used for pre-training included supervised learning for sex classification, and contrastive learning supervised with age as the proxy meta-data. In line with [20] for contrastive pre-training, we used flip, gaussian blurring, gaussian noise, cutout, and random cropping as the transformations for the pretext task. We also experimented with self-supervised learning via masked image reconstruction by masking up to 90% of the T1-w MRI scans from the UK Biobank dataset.

We fine-tuned end-to-end and evaluated our models' performance on three unique neuroimaging tasks - AD classification, Brain Age Prediction, and PD classification. The datasets for each of these tasks are described in the next section.

## 3 Experimental Setup

### 3.1 Data

In line with similar studies [4], all T1-w brain MRI scans were pre-processed via standard steps for neuroimaging analyses, including: N4 bias field correction [22], 'skull-stripping', linear registration to a template with 9 degrees of freedom, and isotropic resampling of voxels to 2-mm resolution. All images were z-transformed to stabilize model training. The T1-w scans were re-sized to 80 voxels across all dimensions except for the masked image experiments where we used an image size of 96.

The in-domain pre-training datasets included 3D T1-w MRI scans from the UK Biobank (UKB) [21]; the Alzheimer's Disease Neuroimaging Initiative (ADNI) [24] - from 3 phases of ADNI - ADNI1, GO, 2 was used for self-pretraining; The out-of-domain datasets illustrated in Figure 1 include 306,245 separate videos belonging to one of 400 different human actions from the Kinetics-400; 3D synthetic T1-w MRIs generated using a Latent Diffusion model (LDM) trained on UKBB [17]. Further, we included the multi-site 'Big Healthy Brains' (BHB) dataset with 3D T1-w MRI 10,420 scans from 7,764 subjects, aggregated from 13 sites (mean age = 32.0 (19.0) years) [7].

For AD classification, ADNI was respectively split into 2,577, 302, and 1,219 T1-w scans into training, validation, and test sets, with unique subjects restricted to a specific fold, thus avoiding data leakage given the longitudinal data. The T1-w scans with valid gray matter maps from ADNI (4096 scans from 1188 subjects, 55.9-94.4 years, 1932F/2164M) were used to fine-tune the BHB dataset [7]. We used the Open Access Series of Imaging Studies (OASIS) [12] as an out-of-distribution test dataset. For brain age prediction, we analyzed only control T1-w MRIs from ADNI (2394 scans from 585 subjects, 60.4-94.4 years, 1189F/1205M), the scans were respectively split into 1511, 183, 700 for training, validation and testing and OASIS (417 scans, 43.5-95.6 years, 253F/164M) for testing. For PD classification we analyzed data from two private datasets, (1) 467 scans from Chang Gung University, Taiwan (Taiwan), split into 378, 42, 47 for training, validation, and testing and (2) 164 scans from University of Pennsylvania (UPenn) as a test set. See **Table 1** for a summary of the in-domain datasets.

**Table 1.** Description of brain MRI data included in this work along with the 'Big Healthy Brains' (BHB)[3] dataset with 3D T1-w MRI 10,420 scans from 7,764 subjects, aggregated from 13 sites (mean age = 32.0 (19.0) years).

| Dataset | N | #Subjects | Age (years) |
|---------|---|-----------|-------------|
| LDM   | 100,000 | 100,000 | 44.0-82.0 |
| UKB   | 38,703  | 38,703  | 44.6-82.8 |
| ADNI  | 4,098   | 1,188   | 55.7-92.8 |
| OASIS | 600     | 600     | 43.5-97.0 |
| Taiwan| 467     | 467     | 20.0-80.0 |
| UPenn | 164     | 164     | 50.0-86.0 |

### 3.2 Model Training and Testing

We performed a random search to select hyperparameter values, including the learning rate {2e-3 to 1e-5}, optimizer {SGD, ADAM, ADAM with weight decay}, weight decay between 0 and 0.2, batch size {1, 4, 8, 16}, and attention heads {4, 8, 12}, encoder layers {3, 6, 8} for the NiT, MiNiT. All models were usually trained for up to 100 epochs with an early stopping rate of 30 epochs. Models were usually fine-tuned for 30 epochs. We experimented with a cosine learning rate scheduler and learning rate warmups up to 10 epochs [20] to improve training convergence. We used binary cross-entropy loss for all classification tasks, L1 loss for the brain age prediction task, and for masked image reconstruction in line with [26], we used a weakly supervised contrastive loss [7] for self-supervised pre-training of the CNNs. We experimented with different image augmentations during training and fine-tuning.

We evaluated our models with the receiver-operator characteristic curve-area under the curve (ROC-AUC), and measures including accuracy, sensitivity and specificity based on a threshold optimized using Youden's Index [27]. In our results, we present an average over two subsequent runs for each experiment. We also tested our models on out-of-distribution (OOD) test sets in a zero-shot setting without any additional fine-tuning.

### 3.3 Model Interpretability

We conducted an occlusion sensitivity [28] analysis to visualize the salient brain regions used by our models. Occlusion involves systematic masking of brain regions with a black patch and aggregating the CNN predictions with and without the occlusion, to create a heatmap.

## 4 Results

**Table 2** summarizes the performance of different configurations of vision backbones and initializations we investigated. We evaluated our models on three neuroimaging tasks with unseen test data, as shown in **Table 3** and **Figure 2**.

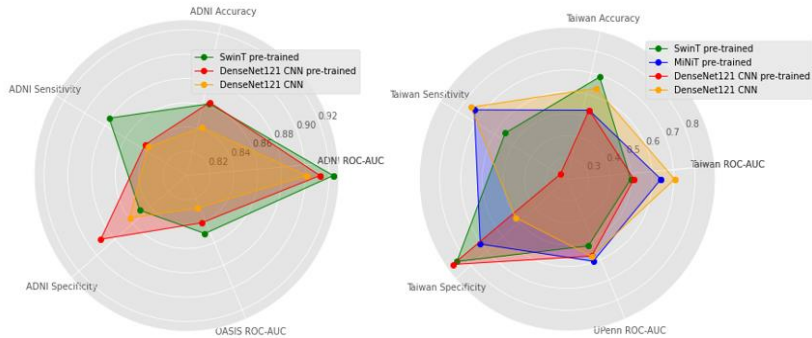

**Figure 2.** Spider plots illustrates test performance metrics of top training configurations across multiple tasks, Left - AD classification, Right - PD classification. This corresponds to Table 3.

6**Table 2.** AD classification using different training configurations.

| Vision Backbone | Initial | ADNI (fine-tuned) | | | | OASIS (0-shot) ROC-AUC↑ |
|---|---|---|---|---|---|---|
| | | ROC-AUC↑ | Accuracy↑ | Sensitivity↑ | Specificity↑ | |
| Swin-T (27M) | Scratch | **0.849** | 0.801 | 0.723 | 0.859 | **0.833** |
| | Kinetics-400 + supervised | **0.924** | 0.862 | 0.881 | 0.849 | **0.850** |
| ViT B/16 scaled down (8M) | Scratch | 0.850 | 0.779 | 0.748 | 0.801 | 0.788 |
| | UKB + masked image | 0.873 | 0.827 | 0.780 | 0.861 | 0.819 |
| MiNiT (3.6M) | Scratch | 0.851 (0.001) | 0.782 (0.022) | 0.768 (0.020) | 0.8189 (0.017) | 0.813 (0.003) |
| | LDM + supervised | 0.844 | 0.783 | 0.773 | 0.790 | 0.790 |
| NiT (2M) | Scratch | 0.805 (0.006) | 0.744 (0.003) | 0.752 (0.003) | 0.738 (0.003) | 0.809 (0.008) |
| 3D Dense-Net121 CNN (11M) | Scratch | **0.901 (0.030)** | 0.842 (0.028) | 0.840 (0.016) | 0.889 (0.001) | **0.828 (0.024)** |
| | UKB + contrastive | **0.912 (0.005)** | 0.863 (0.006) | 0.843 (0.016) | 0.896 (0.020) | **0.841 (0.006)** |
| | ADNI + contrastive | 0.906 (0.003) | 0.838 (0.004) | 0.760 (0.037) | 0.888 (0.037) | 0.818 (0.013) |
| | BHB + contrastive | 0.904 (0.005) | 0.845 (0.005) | 0.786 (0.038) | 0.910 (0.002) | 0.824 (0.023) |
| | LDM + contrastive | 0.892 (0.010) | 0.842 (0.009) | 0.751 (0.023) | 0.889 (0.001) | 0.724 (0.059) |
| 3D Tiny DenseNet CNN (1.8M) | Scratch | **0.897 (0.008)** | 0.832 (0.021) | 0.836 (0.023) | 0.841 (0.045) | **0.838 (0.006)** |

**Table 3.** Adaptation to multiple downstream tasks and generalization to out-of-distribution (OOD) datasets with 0-shot testing. The pre-training N includes: Kinetics-400 (306, 245), LDM (100,000), UKB (38703).

| Vision Backbone | Initial | AD Classification (ROC-AUC) ↑ | | Brain Age Prediction (MAE) ↓ | | PD Classification (ROC-AUC) ↑ | |
|---|---|---|---|---|---|---|---|
| | | ADNI | OASIS (0-shot) | ADNI | OASIS (0-shot) | PD dataset A | PD dataset B (0-shot) |
| SwinT | Scratch | 0.849 | 0.833 | 16.37 | 9.396 | 0.583 | 0.588 |
| | Kinetics-400 + supervised | **0.924** | 0.850 | **4.799** | 9.124 | 0.498 | 0.570 |
| MiNiT | Scratch | 0.851 | 0.813 | 5.169 | 10.291 | 0.595 | 0.555 |
| | LDM + supervised | 0.844 | 0.790 | 5.424 | 11.004 | **0.640** | 0.572 |
| 3D DenseNet 121 CNN | Scratch | 0.901 | 0.828 | 4.837 | 8.306 | 0.513 | 0.596 |
| | UKB + contrastive | **0.912** | 0.841 | **3.582** | 5.701 | **0.704** | 0.5205 |

We conducted an ablation analysis to study the effect of % fine-tuning data used in **Table 4** and **Figure 3**. For model interpretability we conducted an occlusion-based sensitivity analysis for the AD and PD classification, as shown in **Figure 4.**

**Table 4.** Tabulation of the ablation study illustrated in Figure 2. Description of the test performance of different training configurations using 10% and 100% of ADNI training data.

| Backbone | Initialization | Test ROC AUC with 100% training data from ADNI | Test ROC AUC with 10% training data from ADNI |
|---|---|---|---|
| SwinT | Scratch | 0.849 | 0.677 |
|  | Kinetics-400 | **0.924** | **0.759** |
| MiNiT | Scratch | 0.851 (0.001) | 0.772 (0.003) |
|  | LDM + supervised | 0.844 | **0.809** |
| 3D DenseNet 121 CNN | Scratch | **0.901 (0.030)** | **0.858 (0.038)** |
|  | UKB + contrastive | **0.912 (0.005)** | **0.872 (0.007)** |
|  | ADNI + contrastive | 0.906 (0.003) | 0.744 (0.084) |
|  | BHB10K + contrastive | 0.904 (0.005) | 0.855 (0.006) |
|  | LDM + contrastive | 0.892 (0.010) | 0.724 (0.002) |
| 3D Tiny DenseNet CNN | Scratch | **0.897 (0.008)** | **0.853 (0.010)** |

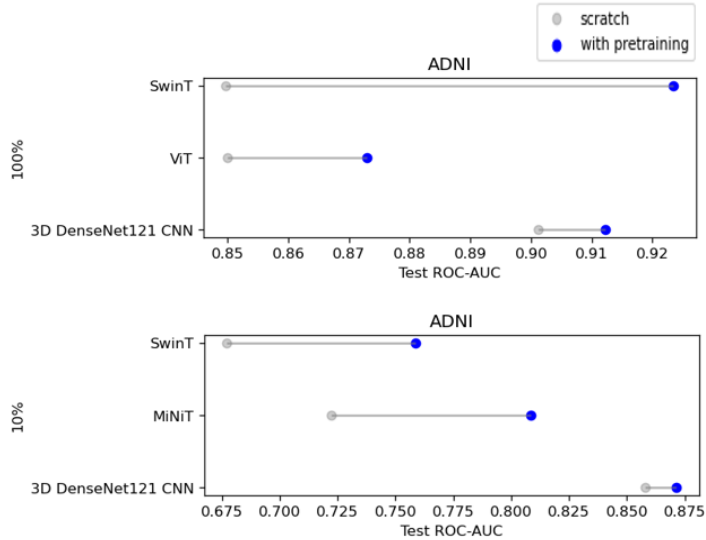

**Figure 3.** Ablation Study: Effect of training data % on test performance using different training configurations. Horizontal line plot shows boost in AD classification performance with *Top*: 100% and *Bottom*: 10% of ADNI training data, the largest fine-tuning dataset in this study.

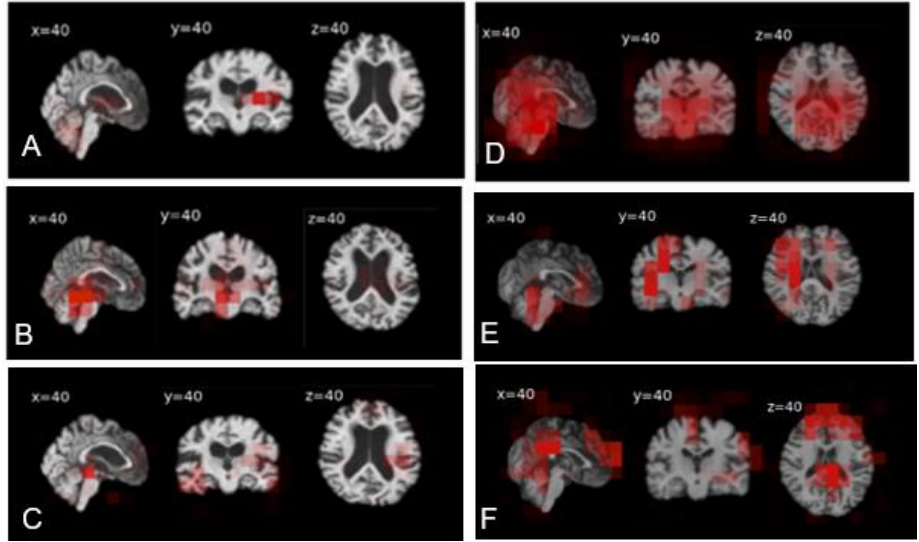

**Figure 4.** Occlusion Sensitivity Analysis for AD (*left column*) and PD (*right column*) classification. *Top* - CNN trained from scratch, *Middle* - Fine-tuned CNN pre-trained on UKB using Contrastive Learning, *Bottom* - Pre-trained SwinT (C) and Pre-trained MiNiT (F). Subcortical areas feature more prominently for classifying PD, where basal ganglia atrophy is common. Attention to the cerebellum may be due to the widened CSF and supratentorial spaces that can accompany brain atrophy.

## 5    Discussion

**Pre-training improved performance across all tasks:** We found that pre-trained backbones consistently outperformed their counterparts trained from scratch across multiple brain MRI tasks. The sampled tasks were of varying difficulty. Minimal fine-tuning of a common pre-trained backbone - for both ViTs and CNNs - allowed them to adapt to unseen tasks. The datasets for each of the tasks also ranged in scale. ViTs benefited more from pre-training using self-supervised and supervised methods relative to CNNs, as seen in **Figure 3**. We observed that self-pretraining on the target dataset (ADNI) using a weakly supervised flavor of contrastive learning helped to boost the CNNs' performance for AD classification. In addition, pre-training with real MRIs from UKB offered the largest performance gain to the 3D CNN across the tasks.

**Emerging pre-training data boosted performance of ViTs**: We demonstrated a boost in performance of ViTs when pre-trained on 3D emerging data including video data and synthetic brain MRIs generated by latent diffusion models. These newer datasets provide an opportunity to scale up the training data for use with current state-of-art vision architectures and overcome limitations in available data in the neuroim-aging/medical domain. As noted by their developers, synthetic data could alleviate privacy concerns associated with sensitive patient data, or any personal data.

**CNNs were robust in limited data scenarios**: We explored the effect of the training data size on the vision backbones and the associated pre-training. We identified the robustness of CNNs in limited data settings. Even when training with only 10% of the training data from ADNI, CNNs saw a relatively lower drop in test performance as presented in **Table 4**. CNNs also benefited with a boost in performance when trained from pre-trained initialization on in-domain MRI data.

**Pre-training improved generalization to out-of-distribution datasets and sites:** We observed that a good pre-trained model initialization improved 0-shot generalization without any additional fine-tuning in 5 out of the 9 experiments on unseen sites and datasets. This was shown empirically in **Table 3** for both ViT and CNN backbones in our experiments across all three tasks.

Although we constrained this work to the T1-w MRI modality, we studied it thoroughly given its widespread availability. In future work we plan to train on multiple data modalities to learn complementary features. Further, in recent times there have been innovations in fine-tuning approaches i.e., by fine-tuning only a fraction of the model parameters via adaptors or prompt-based tuning. We plan to extend this work by experimenting on efficient fine-tuning methods.

## 6 Conclusion

We demonstrated the impact of efficiently fine-tuning state-of-the-art 3D vision backbones with a variety of pre-trained initializations for neuroimaging research. We studied multiple brain MRI analysis tasks and datasets pertaining to AD classification, brain age prediction and PD classification, and evaluated the benefits of pre-training. We show the effects of using large-scale video data and synthetic MRI datasets for pre-training vision transformers, and the robustness of CNNs given limited training data.

1010